%% file: paper-arxiv.tex
\documentclass[a4paper,11pt]{article}
\usepackage{epsfig, amsfonts, amsmath, amssymb,fullpage}
\usepackage[lined,boxed]{algorithm2e} 

\newcommand{\GG}{\mathcal{G}}
\newcommand{\VV}{\mathcal{V}}
\newcommand{\EE}{\mathcal{E}}
\newcommand{\cc}{\textsf{c}}
\newcommand{\un}{\textsf{un}}

\newtheorem{defn}{Definition}[section]

\newtheorem{prob}{Problem}[section]

\title{\Large Multiscale approach for the network compression-friendly ordering}
\author{
Ilya Safro\thanks{Mathematics and Computer Science Division, Argonne National Laboratory, {\tt safro@mcs.anl.gov}}
\and Boris Temkin\thanks{Faculty of Mathematics and Computer Science, The Weizmann Institute of Science, {\tt boris.temkin@gmail.com}}
}

\date{}
\begin{document}

\maketitle
\begin{abstract}
We present a fast multiscale approach for the network minimum logarithmic arrangement problem. This type of arrangement plays an important role in a network compression and fast node/link access operations. The algorithm is of linear complexity and exhibits good scalability which makes it practical and attractive for using on large-scale instances. Its effectiveness is demonstrated on a large set of real-life networks. These networks with corresponding best-known minimization results are suggested as an open benchmark for a research community to evaluate new methods for this problem.
\end{abstract}

\section{Introduction}
\par Finding a suitable compressed representation of large-scale networks is intensively studied in both practical and theoretical branches of data mining \cite{wojciech-compress,micah-compr,compr-social,boldi-vigna,bollt-compr}. In particular, the success of applying some of the recently proposed compression schemes \cite{compr-social,boldi-vigna,ad-gcbfs-09} strongly depends on the ``compression-friendly'' arrangement of network nodes. Usually, the goal of these arrangements is to order the nodes such that the endpoints of network links (edges) are located as close as possible. Doing so leads to a  more compact representation of links and allows a better performance of compression schemes and network element access operations.

\par In \cite{compr-social}, Chierichetti et al. propose a combinatorial optimization problem, namely, the minimum logarithmic arrangement problem, that seeks a nearly optimal information-theoretical compressed encoding size for all network links. This is achieved by ordering the network nodes and assigning to them unique integer values (ids) such that the endpoints of a link will obtain close values. The problem has been  proven to be NP-hard, and two heuristics for ordering the social networks were given. In this paper, we present a multiscale method for approximating a generalized {\it link-weighted} and {\it node-volumed} version of the minimum logarithmic arrangement for {\it general} networks. The importance of a link-weighted property (when each link is assigned by a nonnegative weight) for this problem is twofold. First, the link weight can measure the significance of that link. For example, in many cases we have  information regarding its access frequency, and we would prefer that frequently accessed links would be compressed better. Second, the multiscale algorithmic framework admits a natural aggregation of {\it weighted} links at different scales of the problem representation (discussed later). Similary to the link weights, all nodes are assigned by nonnegative volumes that represent the length of their segments captured in the ordering of the network nodes.

\par Our multiscale algorithm is based on the algebraic multigrid (AMG) \cite{bmr} methodology for linear ordering problems \cite{survey:petit,safro2005}. The main objective of the AMG-based framework is to construct a hierarchy of problems ({\it coarsening}), each approximating the original problem, but with fewer degrees of freedom. This is achieved by introducing a sequence of successive projections of networks' graph Laplacians into lower-dimensional spaces and solving the problem in them. The multiscale framework has two key advantages that make it attractive for applying on modern large-scale instances: it exhibits a linear complexity, and it can be relatively easily  parallelized and implemented by using standard matrix-vector operations. Another advantage of the multiscale framework is its heterogeneity, expressed in the ability to incorporate external appropriate optimization algorithms (as a refinement) in the framework at different scales. For more detailed surveys on the multiscale methods and their parallelization  for combinatorial optimization problems, we refer the reader to \cite{vlsicad,walshaw-inbook,boman-advances}.

\par In contrast to \cite{compr-social}, the main goal of this work is to provide a {\it generic} solver for the minimum logarithmic arrangement for general networks. Social networks (including big parts of Web graph) commonly consist of a combination of structural properties (such as degree distribution, small diameter and expander-like topology) that enable fast greedy methods (usually based on some preferential ordering of graph traversal) to find arrangements for further  high quality compression. These methods can be successful because most of the links exhibit  good locality \cite{soc-net-evol,soc-locality} and only a small number of them can be considered as global edges that connect well-separated nodes/regions of a network. Thus, these networks can hardly be considered as strongly irregular and difficult instances. In real life, however, there occur many situations when the structure of a stored network (or a part of it) is  complex and irregular (such as decision/detailed supply/infection spread networks and even parts of social networks that do not exhibit power law degree distribution), which creates many contradictions between greedy local decisions and solutions that consider a more global picture. We address this type of problem by introducing a multiscale solver.

\par The experimental part of this work shows how far are the existing state-of-the-art ordering heuristics from being optimal. We demonstrate significant improvement for minimization of logarithmic arrangement for various families of networks (including social networks and other (ir)regular instances). The comparison was performed with several recently introduced methods in  \cite{webgr:impl,boldi-vigna,boldi-permuting,compr-social}. In almost all cases, our solver exhibited better numerical results than previous best-known results.

\par We introduce notation and necessary definitions in Section \ref{prob-def}. The main algorithm is described in Section \ref{sec:alg}. The computational results and discussion about different parameter settings are presented in Section \ref{sec:res}. In Section Section \ref{sec:concl} we conclude and provide possible future research directions. For readers  interested in their own implementation, we note that some details regarding the multiscale scheme (that are not related strongly to the discussed problem) are omitted in this paper; they can be found in \cite{safro2003,safro2005,safro2010,vlsicad}.

\section{Notation and problem definition}\label{prob-def}
\par A network is described by a weighted directed graph $G=(V,~E)$, where $V=\{1,2,...,n\}$
is the set of nodes (vertices) and $E$ is the set of directed edges. If $ij\in E$, then there exists an edge $i \rightarrow j$. Denote by $w_{ij}$ the non-negative weight of the directed edge $ij$ between nodes $i$ and $j$; if $ij\notin E$, then $w_{ij}=0$. Let $\pi$ be a bijection
\[
\pi : ~ V ~ \longrightarrow ~ \{1,2,...,n\}~~~.
\]
The purpose of the link-weighted version of the minimum logarithmic arrangement problem (MLogA) is to minimize
\begin{equation}
 \sum_{ij\in E}w_{ij}\lg|\pi(i)-\pi(j)|
\end{equation}
over all possible permutations $\pi$. The base of a logarithm ($\lg$) will be always 2.
 We define the generalized form of this problem (GMLogA) that emerges during the multilevel solver; each vertex $i$ is assigned with a $volume$ (or $length$), denoted $v_i$. Given the vector of all volumes, $v$, the task now is to minimize the cost
 \begin{equation}\label{probeq}
 \cc (G,x)=\sum_{ij\in E}w_{ij}\lg|x_i-x_j|
 \end{equation}
over all possible $\pi$, where \mbox{$x_i=\frac{v_i}{2}+\sum_{k, \pi(k)<\pi(i)}v_k$}; that is, each vertex is positioned at its center of mass, capturing a segment on the real axis that equals its length. The original form of the problem is the special case where all the volumes are equal.
\par We denote the process of creating a weighted undirected graph $\GG=(\VV,\EE)$ by $\GG \leftarrow \un (G)$, where $G$ is a given directed graph. By applying $\un (G)$, we obtain $\VV\leftarrow V$ and
\[
ij\in\EE \Leftrightarrow ij\in E \text{ or } ji\in E~~~.
\]
The edge weights are accumulated from the set of directed edges
\[
\forall ~ ij\in \EE ~ w_{ij} = \sum_{ij\in E} w_{ij} + \sum_{ji\in E} w_{ji}
\]
allowing use of multigraphs. Both MLogA and GMLogA are formulated for the undirected graphs similarly to their directed versions.
In this paper we will find approximate solutions for the following problem:
\begin{prob}\label{probdef}
Given a directed weighted network graph $G$, the objective is to find a permutation $\pi$ that minimizes $\cc(G,x_{\pi})$, where $x_{\pi}$ is a vector of the respective node coordinates restricted by $\pi$.
\end{prob}
Note that the problem is formulated for both directed and undirected graphs. We define\footnote{$\beta$ with no subindexes will be used where appropriate.} $\beta_{G,x_{\pi}}$ as the main unit for the empirical comparison in Section \ref{sec:res}:
\begin{equation}\label{eq:bpl}
\beta_{G,x_{\pi}} = \cc(G,x_{\pi})/\sum_{ij\in E} w_{ij}~.
\end{equation}
\section{The algorithm}\label{sec:alg}
\par
\begin{figure}
 \centering
 \includegraphics[width=7cm,bb=0 0 611 486]{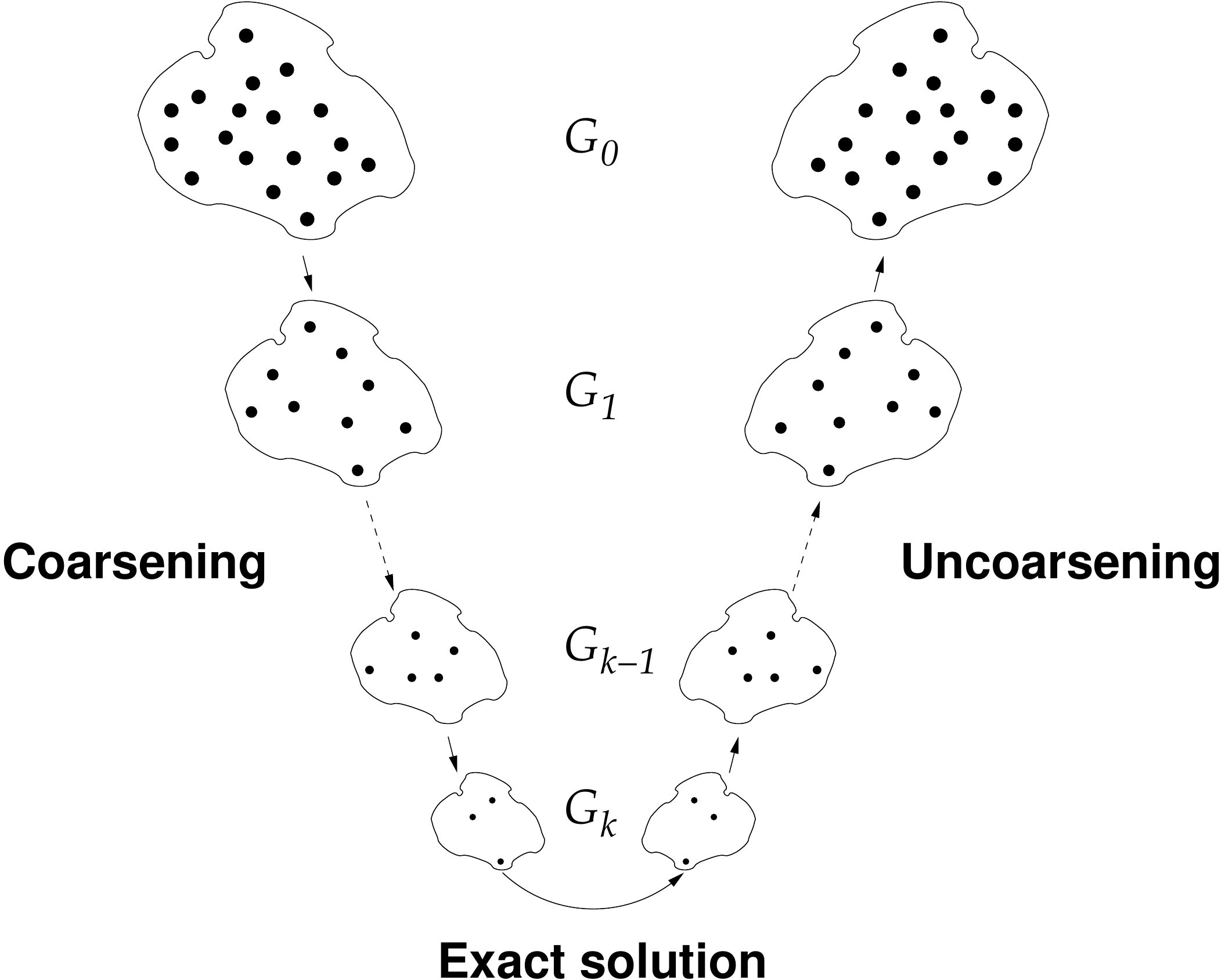}
 \caption{Scheme of V-cycle. The arrows show the order in which smaller graphs are created and revised for approximation.}
 \label{fig:v-cycle}
\end{figure}

 In the multiscale framework we construct a hierarchy of decreasing-size graphs,  $G_0,G_1,...,G_k$, by {\it coarsening}, starting from the given graph $G_0 \leftarrow \un (G)$. In other words, we create an undirected version of a given directed graph by assigning to the undirected edges bigger weights if, in the original unweighted graph, the directed edges had been  reciprocal. At the coarsest level Problem \ref{probdef} is solved exactly, and starting from the $(k-1)$th level it is formulated and approximated by successive prolongation of the solution from the previous coarser level.  This entire process is called a $V$-$cycle$ (see Figure \ref{fig:v-cycle}).

\subsection{Coarsening}
\input{coarsening}
\subsection{Uncoarsening}
The uncoarsening process starts by solving Problem \ref{probdef} at the coarsest level.
Since the number of nodes at the coarsest level is very small (in our tests it was 9),
the problem can be solved exactly by exhaustive search.

\subsubsection{Minimizing the contribution of one node}\label{sec:one-node}
\input{one_node}

\subsubsection{Initialization}\label{sec:init}
Given is the arrangement of the coarse-level aggregates in its generalized form, where the center of mass of each aggregate $j\in C$ is positioned at $x_{I(j)}$ along the real axis. We initialize  the fine-level arrangement by letting each seed $j\in C$ inherit the position of its respective aggregate: $x_j \leftarrow x_{I(j)}$. However, in contrast to the {\it initialization by stages} (that was done for the minimum $p$-sum and the minimum workbound problems \cite{safro2005}) in which the $F$-nodes' positions are calculated according to {\it all} already initialized fine-level neighbors, we found that the initialization that is inspired by the principles of the classical AMG coarse solution projection $x^f ~ \leftarrow ~ P^T x^c$ can produce significantly better approximations. In other words, after initializing the fine-level $C$-nodes, $F$-nodes will be placed one by one using Algorithm \ref{linDensityAlgo}, which takes into account only the $C$-neighbors of a current $F$-point. The entire initialization scheme is presented in Algorithm \ref{alg:init}.
\begin{algorithm}
  \caption{Initialization of the fine level}
  \label{alg:init}
  \KwIn{arrangement of a coarse graph}
  \KwOut{initial ordering of fine level nodes}

  \tcc{Initialize $C$-points}
  \For {$\text{all } j\in C$} {
    $x_j \leftarrow x_{I(j)}$
  }
  \tcc{Initialize $F$-points}
  \For {$\text{all } j\in F$} {
    $x_j\leftarrow$ find best coordinate for $j$ using Algoriothm \ref{linDensityAlgo} with $N_j$
  }
  \tcc{Legalization of the coordinates}
  \For {$\text{all } j\in V$} {
    $x_i = v_i/2 + \sum_{x_k<x_i}v_k$
  }
\end{algorithm}

\subsubsection{Relaxations and strict minimization}\label{sec:relax}
\par The initialization produces a feasible (that satisfies the constraint of (\ref{probeq})) solution inherited from the coarse level. This solution is further improved by employing a small number of relaxation sweeps of two types: {\it compatible} and {\it Gauss-Seidel} (GS). The goal of the compatible relaxation is to improve the positions of $F$-nodes while keeping the $C$-nodes invariant. One sweep of the compatible relaxation can be described by the two last ``for'' cycles of Algorithm \ref{alg:init} when in the first cycle the best coordinate is calculated over {\it all} neighbors.
\par Having the improved ordering of $F$-nodes relative to the $C$-nodes, we apply GS relaxation. This relaxation traverses {\it all} nodes one by one and tries to find a best position for every node such that its own contribution to the total sum of logarithms will be minimized. In all our experiments we observed that, after improving the order with the compatible relaxation, the GS relaxation has almost no influence on the order. This is in contrast to other multiscale linear ordering methods in which the GS relaxation was one of the most powerful and crucial components.

\subsubsection{Strict minimization}\label{sec:strict}
\par We applied two methods of strict minimization (SM, also known as ``local refinement''): node-by-node minimization (N-N) and window minimization (WM). The principal difference between relaxations and SM is that in SM each change is accepted if and only if it reduces the entire sum (\ref{probeq}) and not the local node contribution only (\ref{generalEnergy}).
\par In the N-N minimization all nodes are scanned according to their current order, and each vertex $i$, in its turn, is checked for the best position over some small enough segment that $i$ belongs to. The $k$ left and $k$ right candidate positions are scanned, and the one with the minimal cost is chosen.
\par In the WM (see \cite{safro2005}), the total cost of the arrangement is reduced by numerical minimization of a collective contribution of a small group of consecutive nodes. Given a current feasible solution $\tilde{x}$ of the graph logarithmic arrangement, denote by $\delta_i$ a small correction to $\tilde{x}_i$. Denote by  $\mathfrak{W}=\{i_1=\pi^{-1}(s+1),...,i_q=\pi^{-1}(s+q)\}$ a subset of successive $q$ vertices in the current arrangement. The goal of the local minimizaiton problem is then to find $\delta$ such that
\begin{equation}\label{quadratic}
 \sum_{\substack{i,j \in \mathfrak{W}}}w_{ij}\lg|\tilde{x_i}+\delta_i-\tilde{x_j}-\delta_j| + \sum_{\substack{i \in \mathfrak{W}\\j\not\in \mathfrak{W}}}w_{ij}\lg|\tilde{x_i}+\delta_i-\tilde{x_j}|
\end{equation}
is minimized. The solution to this optimization problem can be approximated numerically by using  techniques similar to \ref{sec:one-node}.
\par The situation with strict minimization is similar to that with GS. In all our experiments, we found no need to use the more complicated WM. The simple N-N minimization with small distance parameter $k<5$ was enough to obtain the best results. This is a major  advantage because in most multiscale schemes the significant part of the running time is spent on the uncoarsening. The lack of need (or, more correctly, optionality) for GS and WM (or other collective optimization) also strongly advocates the choice of the C-nodes  because the further compatible relaxation solves the problem well.
\par After having defined all the components of the V-cycle, we present the full scheme in Algorithm \ref{alg:vcycle}.
\begin{algorithm}
  \caption{ms-GMLogA: Full scheme of one V-cycle for GMLogA.}\label{alg:vcycle}
  \KwIn{$L_f$ is the Laplacian of undirected graph $\GG$ (initially obtained from $\un(G)$)}
  \KwOut{approximated graph minimum logarithmic arrangement}
  \If {$\GG$ is small enough} {
    solve the problem exactly\\
    return the arrangement
  }
  \tcc{Coarsening}
  calculate algebraic distances $\rho_{ij},~\forall ij\in \EE$\\
  identify the set of $C$-points (see (\ref{eq:coarse-cond}))\\
  construct the fine-to-coarse projection $P$ (see (\ref{interpmat}))\\
  $L_c\leftarrow PL_fP^T$\\
  \tcc{Recursive call}
   $\Pi \leftarrow ~$V-cycle($L_c$)\\
  \tcc{Uncoarsening}
  $\pi_1 \leftarrow ~$initialization($\Pi$)\\
  $\pi_2 \leftarrow ~$compatible-relaxation($\pi_1$)\\
  $\pi_3 \leftarrow ~$GS-relaxation($\pi_2$)\\
  $\pi_4 \leftarrow ~$N-N refinement($\pi_3$)\\
  \Return $\pi_4$
\end{algorithm}

\section{Computational results}\label{sec:res}
\par The implementation of Algorithm \ref{alg:vcycle} is based on the multiscale framework used in \cite{safro2005}. The implementation is nonparallel and has not been fully optimized. The results (arrangements and running times) should be considered only qualitatively and can certainly be further improved by more advanced implementation and multiscale techniques.
\par Because of the practical significance of MLogA, we designed an open site \cite{mloga-site} with the benchmark graphs and a set of numerical results. In this site we present the results listed below and invite the scientists who work on this problem to submit interesting new networks, their solutions, and improved arrangements for the existing networks.
\par Our benchmark consists of 100 graphs of different nature and size (most of them are taken from \cite{davis} and \cite{snap}). For these graphs we created their undirected versions (by applying $\un (\cdot)$) and evaluated our algorithms on both the directed and undirected versions. We evaluated two baseline algorithms:
\begin{itemize}
 \item MinLA+N-N: the multiscale solver that finds an approximation of MinLA \cite{safro2003} followed by GMLogA-oriented N-N fast postprocessing (see Section \ref{sec:strict} with $k=10$);
 \item ms-GMLogA: GMLogA solver described in Algorithm \ref{alg:vcycle}.
\end{itemize}
Clearly, MinLA and GMLogA can have different optimal orderings. However, we observed that qualitative MinLA orderings reinforced by appropriate postprocessing can also lead to good  solutions of GMLogA. In spite of the fact that most of the best solutions were  obtained with ms-GMLogA and because the observation was done on a benchmark of large-scale real-life graphs (and there exists intensive research on MinLA \cite{survey:petit,survey:lai}), we decided to present these results too. We also mention that the Fiedler vector-based solutions (even those that try to achieve a good local minimum for MinLA) usually produce significantly fewer  qualitative solutions even with appropriate postprocessing.
\par In our experiments we use a WebGraph framework \cite{boldi-vigna}, which provides a simple way to manage large graphs. In particular, the last version of WebGraph produces random, lexicographical \cite{boldi-vigna}, Gray \cite{boldi-permuting}, double shingle \cite{compr-social}, and LayeredLPA \cite{webgr:impl} orderings for the graphs. The important feature of these orderings is that they are completely endogenous (i.e., determined by the graph itself), contrary to natural ordering, that is, the ordering in which the graph was created. It would seem that the random ordering always has to be significantly worser, than the lexicographic and natural orderings. However, this is not always true, mostly because of the ways the latter have been produced. For example, in highly parallel systems with shared memory, the parallelized graph traversal algorithms can produce arrangements that have been dumped in parallel from different processors. Each processor produces an arrangement of good locality for a small subset of nodes, but the entire arrangement does not possess this property. In particular, in our benchmark we found two graphs whose random ordering was better than the natural and lexicographic ones.
\par
\begin{figure}
\begin{minipage}[b]{0.5\linewidth} 
\centering
\includegraphics[width=7cm]{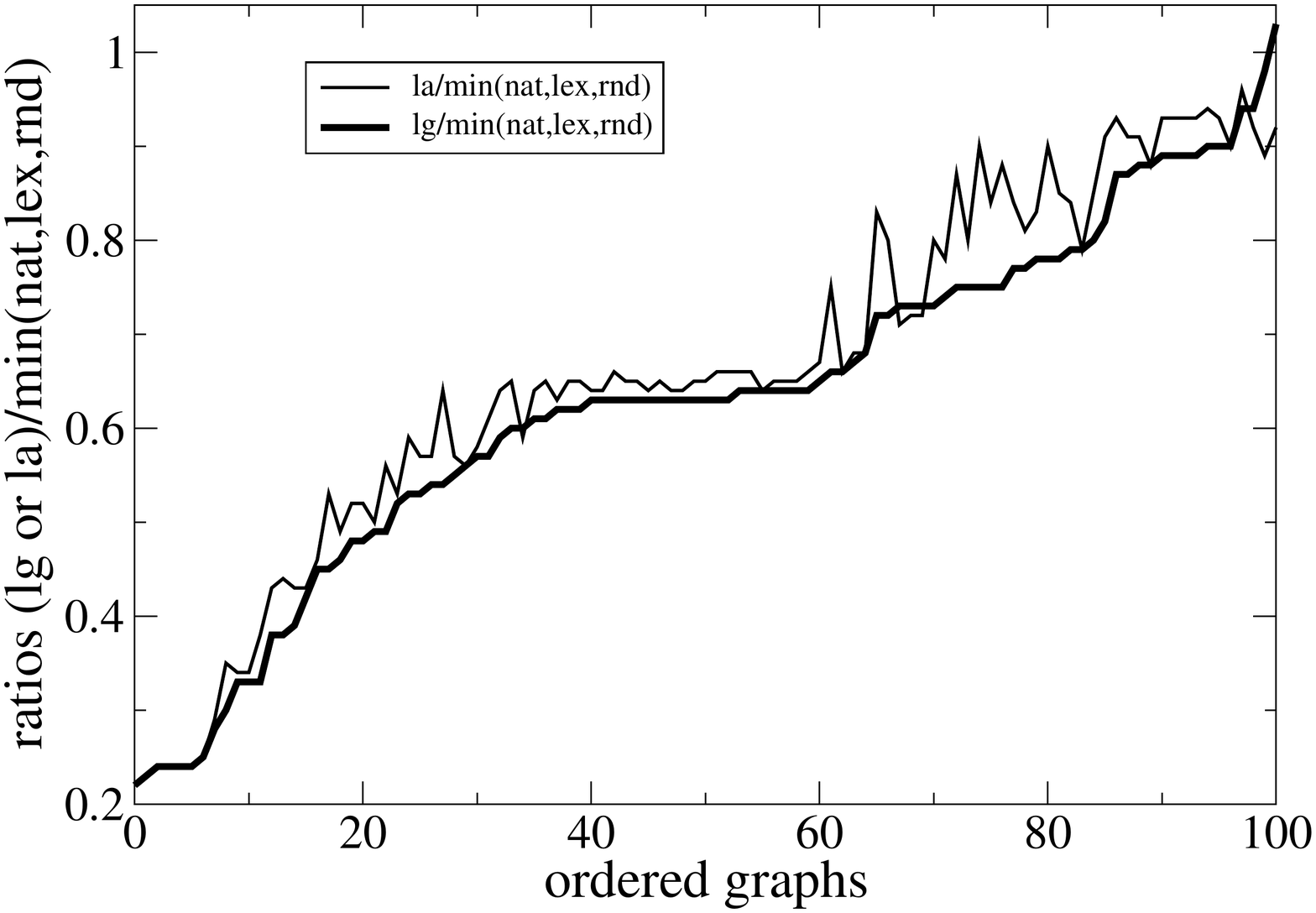}\\
(a) directed\\
\includegraphics[width=7cm]{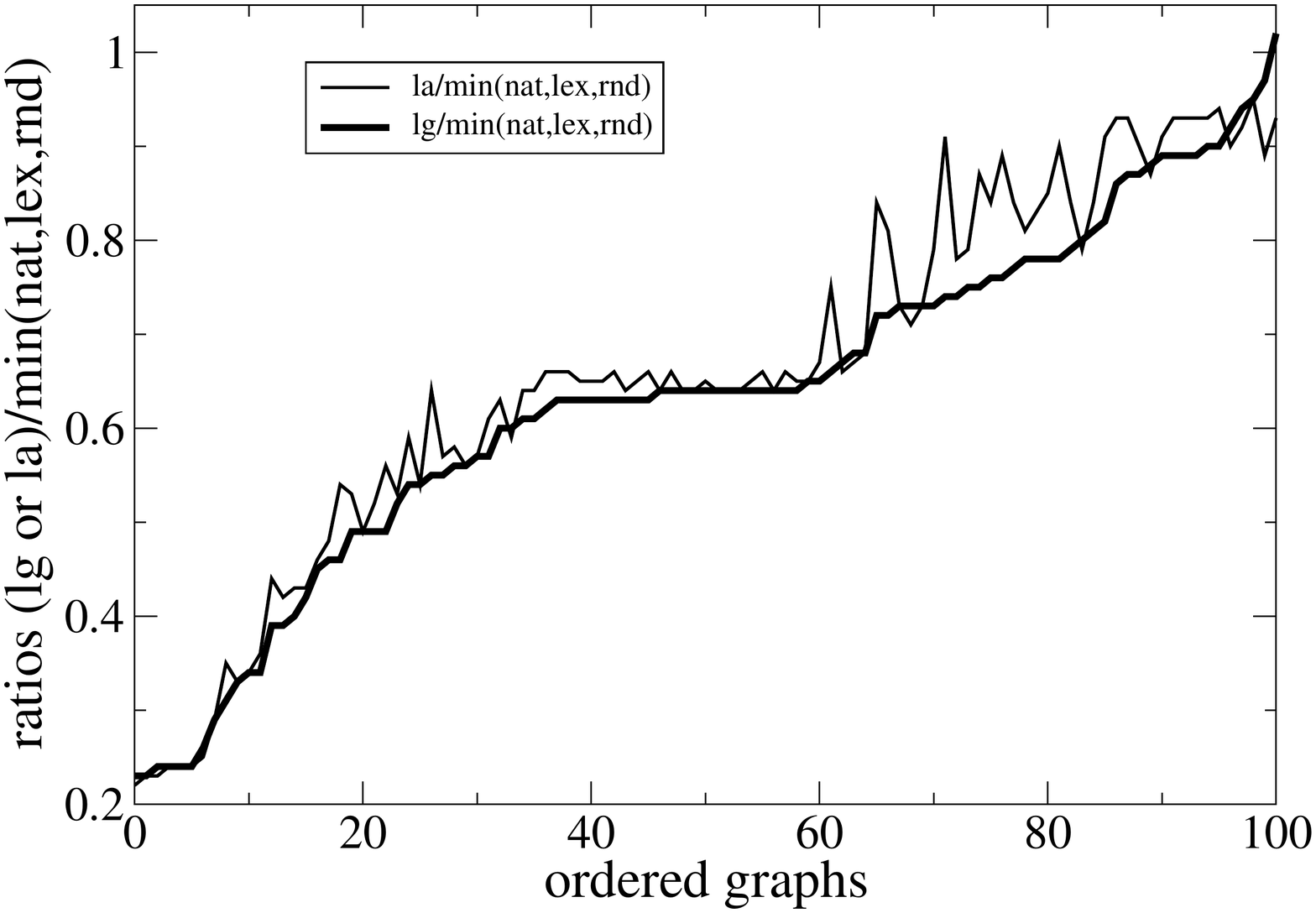}\\
(c) directed\\
\includegraphics[width=7cm]{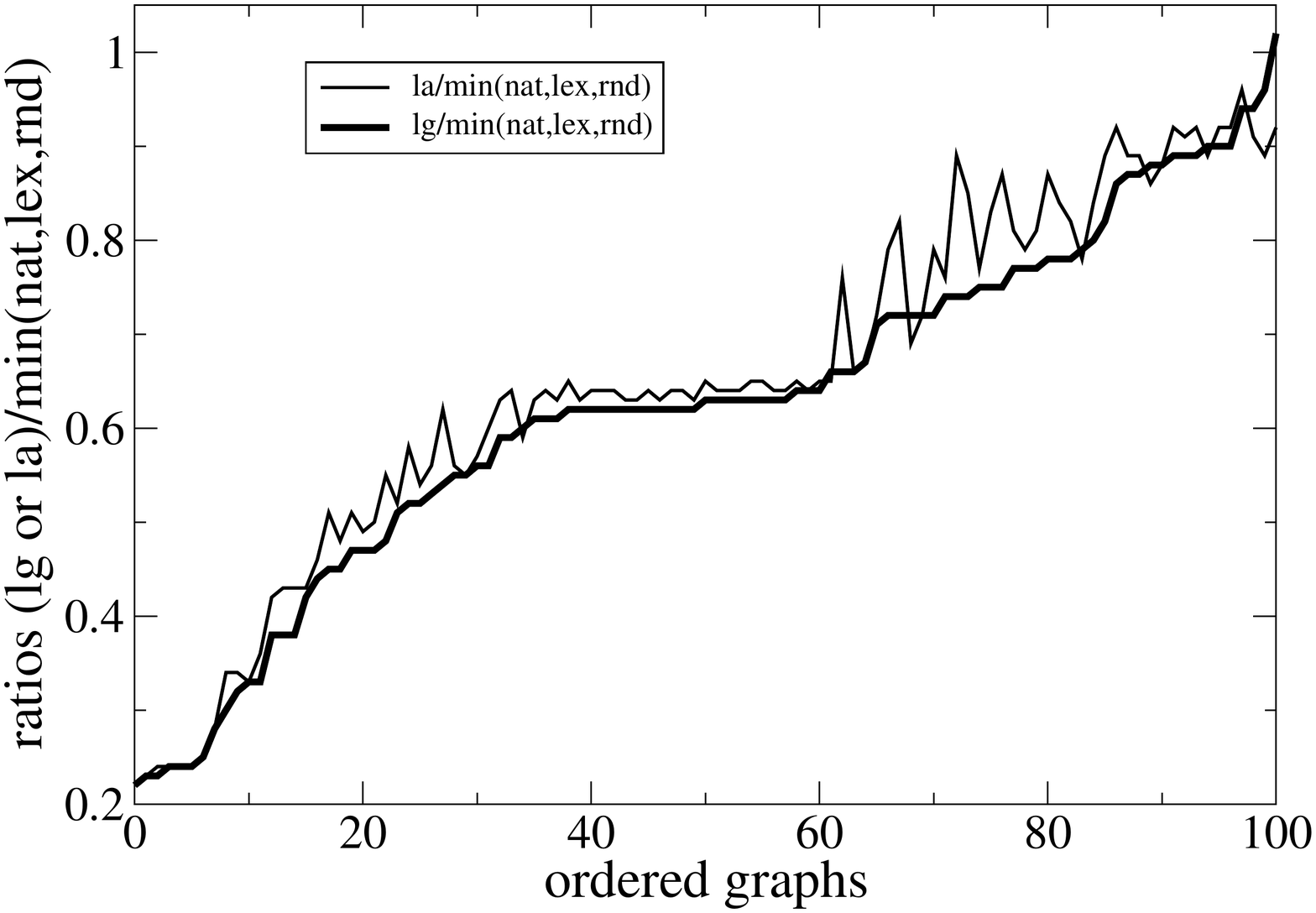}\\
(e) directed
\end{minipage}
\hspace{0.5cm} 
\begin{minipage}[b]{0.5\linewidth}
\centering
\includegraphics[width=7cm]{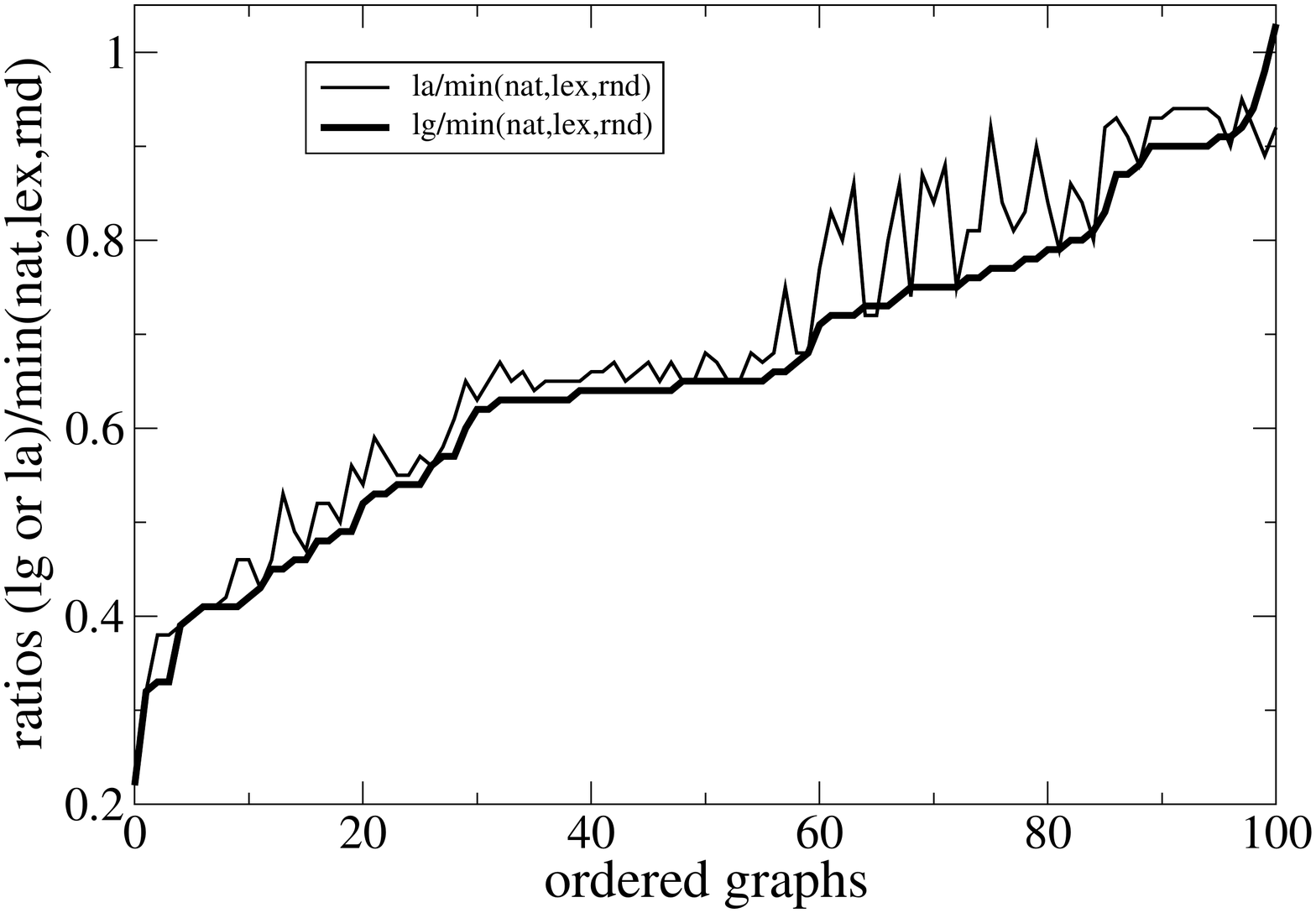}\\
(b) undirected\\
\includegraphics[width=7cm]{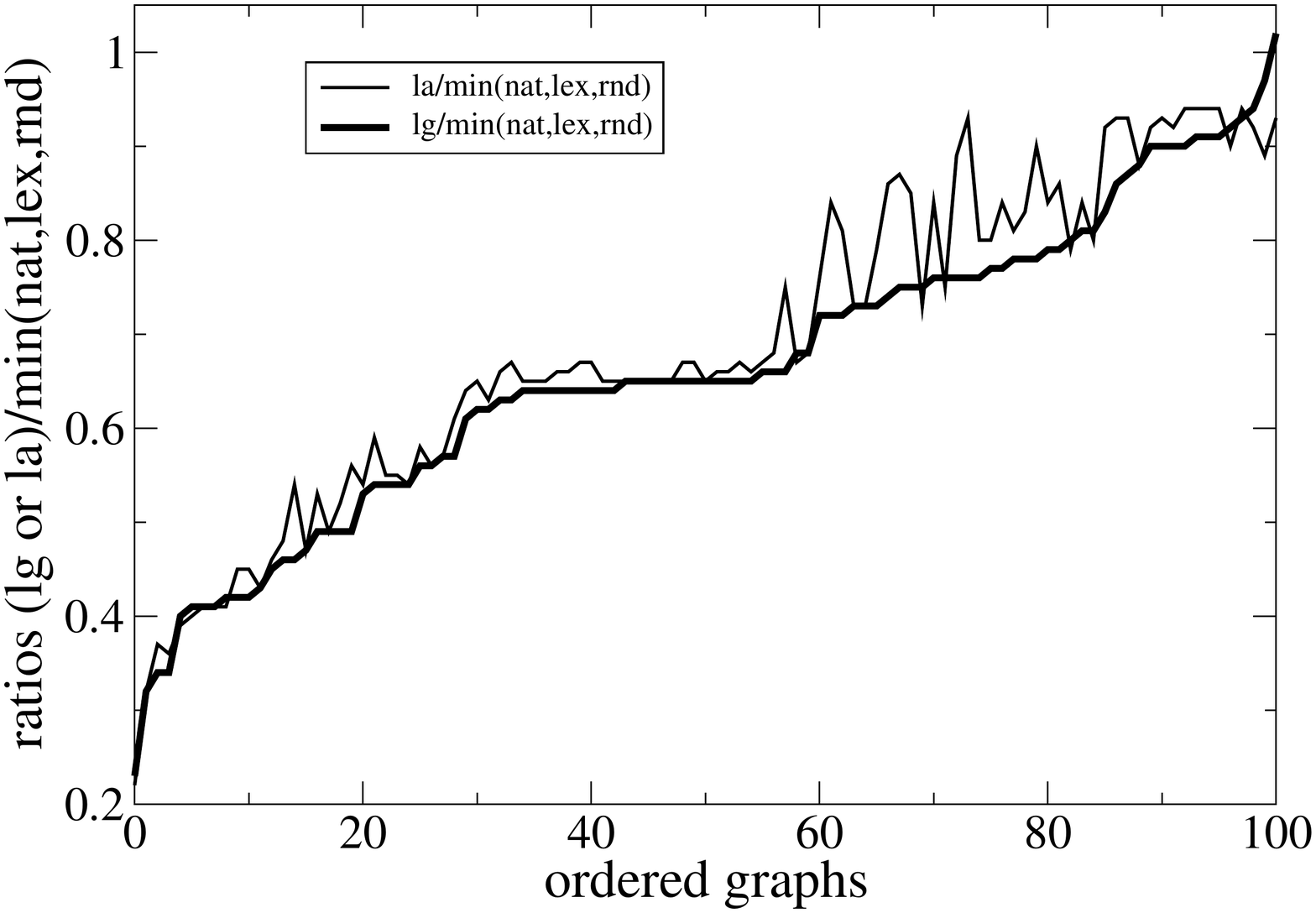}\\
(d) undirected\\
\includegraphics[width=7cm]{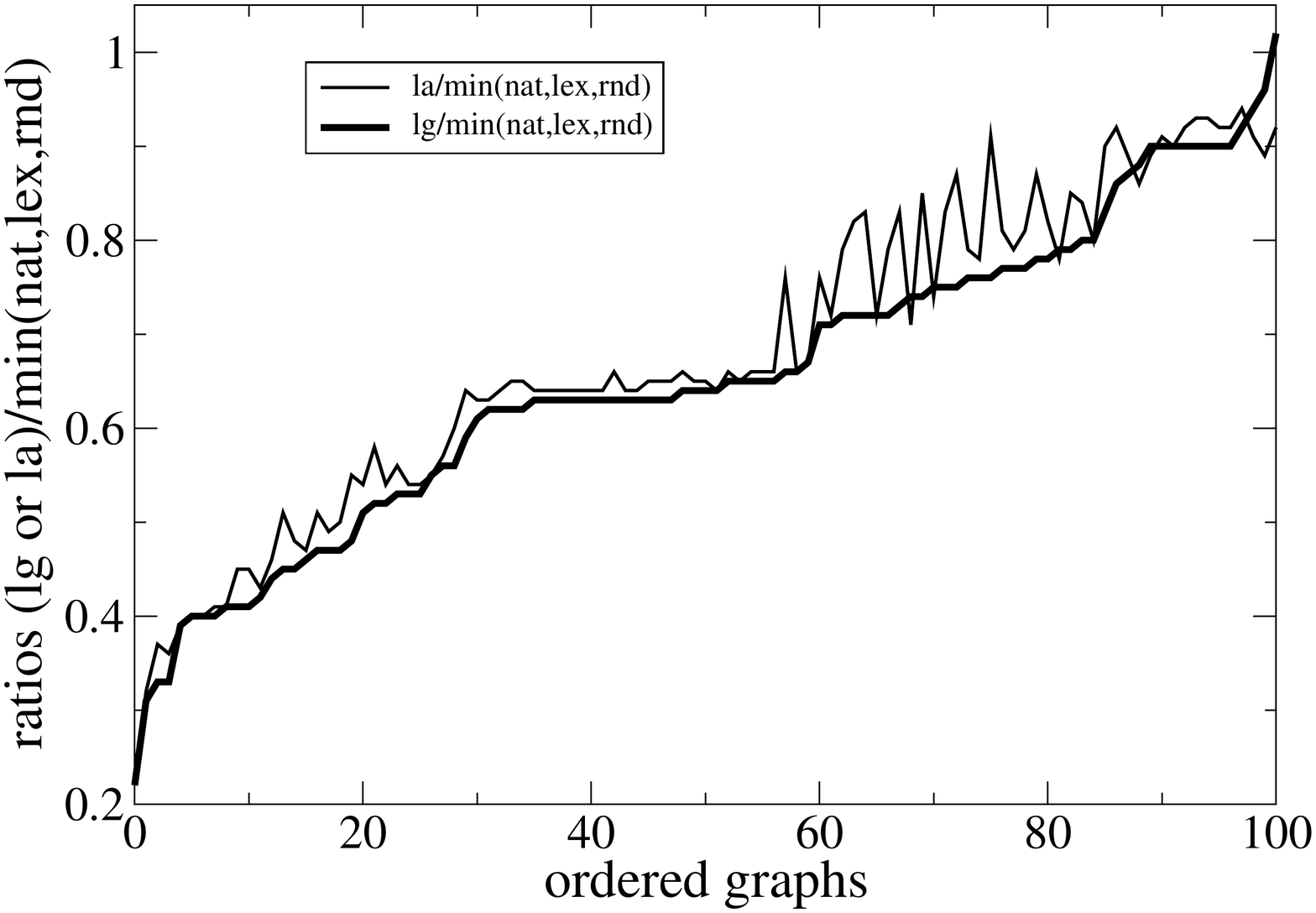}\\
(f) undirected\\
\end{minipage}
\caption{Main comparison results. Notations 'lg' and 'la' correspond to the reults of ms-GMLogA and MinLA+N-N, respectively.}\label{fig:res1}
\end{figure}

\begin{figure}
\begin{minipage}[b]{0.5\linewidth} 
\centering
\includegraphics[width=7cm]{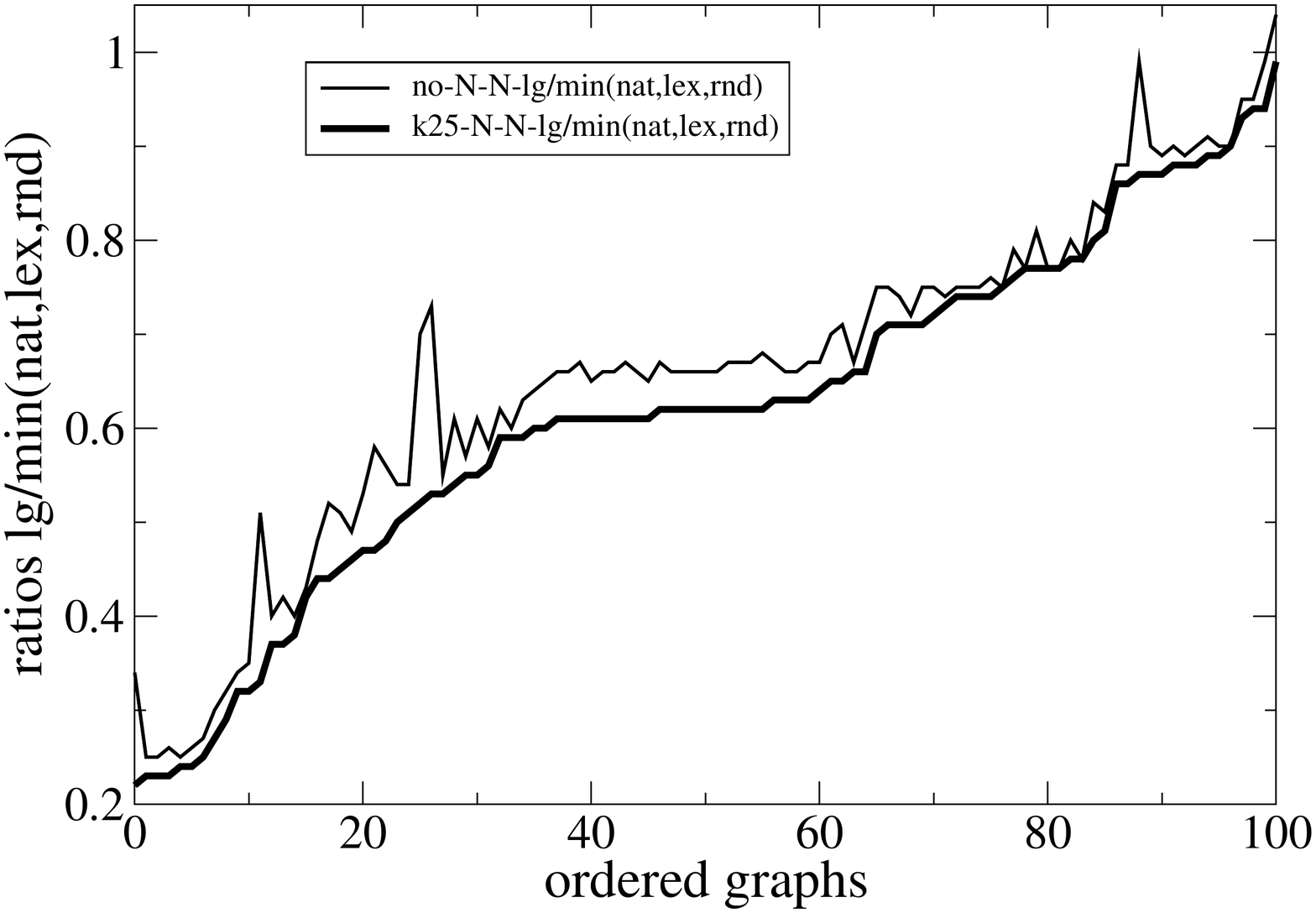}\\
(a) directed
\end{minipage}
\hspace{0.5cm} 
\begin{minipage}[b]{0.5\linewidth}
\centering
\includegraphics[width=7cm]{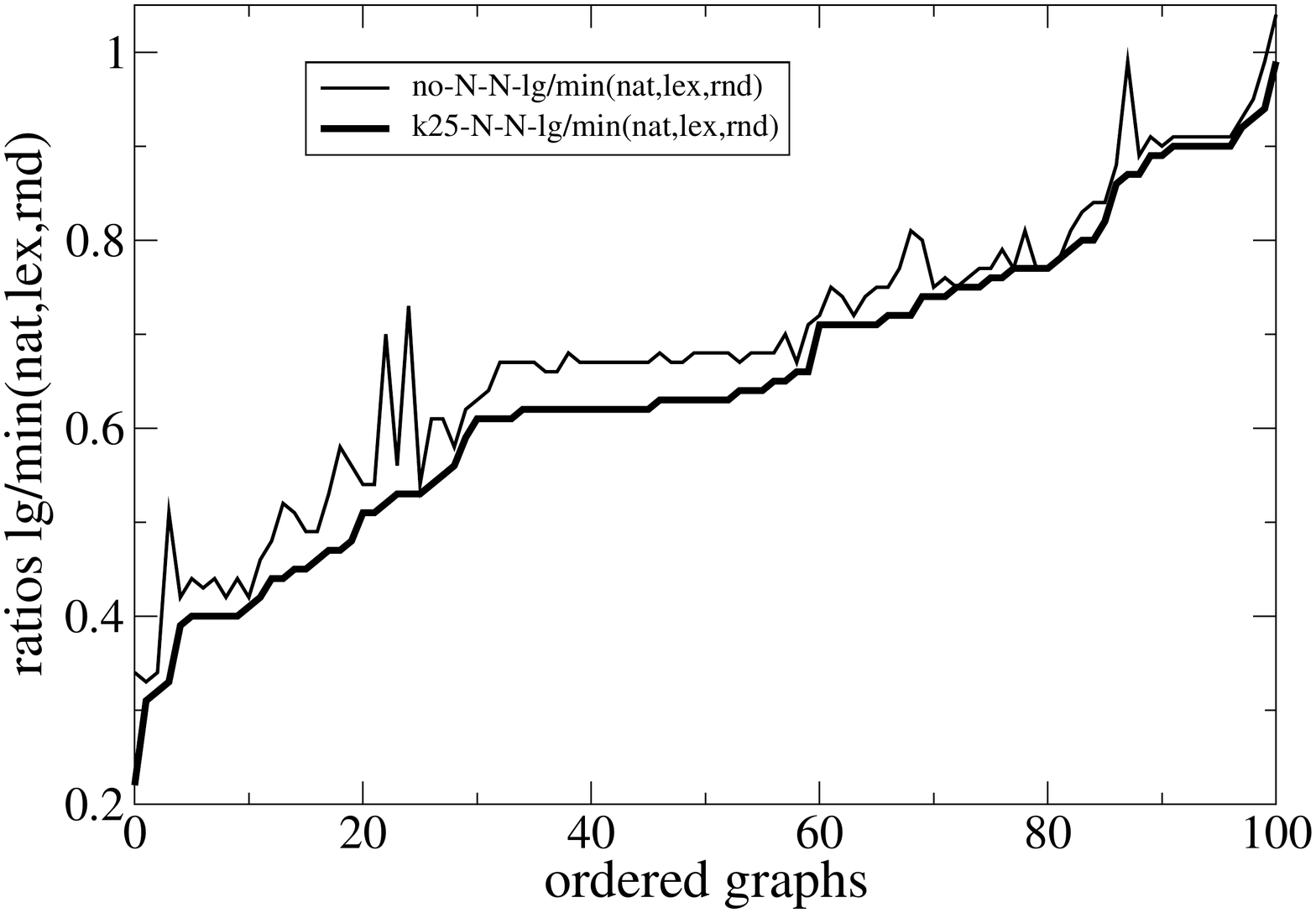}\\
(b) undirected
\end{minipage}
\caption{Comparison of the fastest and the slowest versions of ms-GMLogA. Notations 'no-N-N-lg' and 'k25-N-N'lg' correspond to the fast and slow versions of ms-GMLogA, respectively.}\label{fig:res2}
\end{figure}

 We use the average number of bits per link $\beta$ (\ref{eq:bpl}) as a unit of comparison for the main results presented in Figures \ref{fig:res1}(a-f) and \ref{fig:res2}(a-b). In both figures the left and right columns correspond to the comparison results of directed and undirected graphs, respectively. In Figures \ref{fig:res1}(a-f), each plot area consists of two curves. Each point in bold curves represents a ratio between $\cc(G)$ obtained by ms-GMLogA and $\min(\cc(G, \pi_{nat}), \cc(G,\pi_{lex}), \cc(G,\pi_{rnd}))$  for a particular graph from the benchmark, where $\pi_{nat}$, $\pi_{lex}$, and $\pi_{rnd}$ are the best orderings obtained by natural, lexicographic, and randomized arrangements, respectively. Similarly, each point in regular curves corresponds to the ratio with MinLA+N-N in the nominator. To demonstrate the robustness of the algorithm, we show several sets of its parameters. Each row pair of subfigures corresponds to the same set of parameters for directed and undirected graphs, respectively.
\par In the multiscale algorithms, there are several possible sources (see \cite{vlsicad}) of potentially higher than linear complexity (or linear with coefficients that are too big). Here we address the four most important: (a) the number of iterations in the relaxations/refinements, (b) the complexity of one iteration of relaxation/refinement, (c) the order of interpolation that can increase the complexity of the coarse graphs, and (d) too many C-nodes (addressed in Appendix A). The first set of parameters (Figures \ref{fig:res1}(a,b)) (which is a suggested default set) contains a mild configuration. The order of interpolation (number of nonzero entries in one row of $P$) is only 1; $k$ in N-N refinement is 5; and the number of relaxations of any kind is at most 20. The calculation of the algebraic distance couplings in this set is based on 5 randomly initialized vectors and only 20 iterations of JOR. For almost all graphs, we observed that ms-MLogA produces better orderings than does MinLA+N-N. On average, ms-MLogA improves the graphs by more than 40\% of their ordering cost in comparison to natural, lexicographic, and random arrangements.
\par We note that the most beneficial graphs (those that have ratios from 0.2 to 0.4) in both directed and undirected versions come from different collections such as VLSI design networks, part of Web network, road maps, and Amazon links. In general, these graphs differ significantly  in their structural properties. We observed that no particular part of the benchmark was less beneficial than another, thus attesting to the generality of the proposed method.
\par The difference between improvements obtained with different parameter sets is not significant; however, we provide them to demonstrate the robustness of the method. In the second set of parameters (Figure \ref{fig:res1}(c,d)) the number of random initial vectors in the algebraic distance coupling is reduced to 1 only. It leads to the faster coarsening (while keeping all convergence and model properties of the algebraic distance \cite{chen-safro-algdist-full}) and potentially weaker decisions regarding edge strengths based on the algebraic distance. In the third set of parameters (Figure \ref{fig:res1}(e,f)) we increase $k$ in N-N refinement from 5 to 10 and the interpolation order from 1 to 2 at the finest level of hierarchy. At all coarse levels these parameters are increased too, according to the logarithmic increase scale described in \cite{safro2003}.
\par In Figure \ref{fig:res2}(a,b) we show the difference between the very fast version of ms-GMLogA with no refinement and fast relaxation (5 sweeps) and the slowest version with very aggressive relaxation/refinement parameters ($k=25$ and 40 sweeps). Each subfigure contains two curves: regular and bold. The regular corresponds to the fast version and the bold to the slowest one. It is easy to see that the difference between these two versions is not significant and one can certainly use a fastest version to obtain qualitative results.
\par
\begin{figure}
\begin{minipage}[b]{0.5\linewidth} 
\centering
\includegraphics[width=7cm]{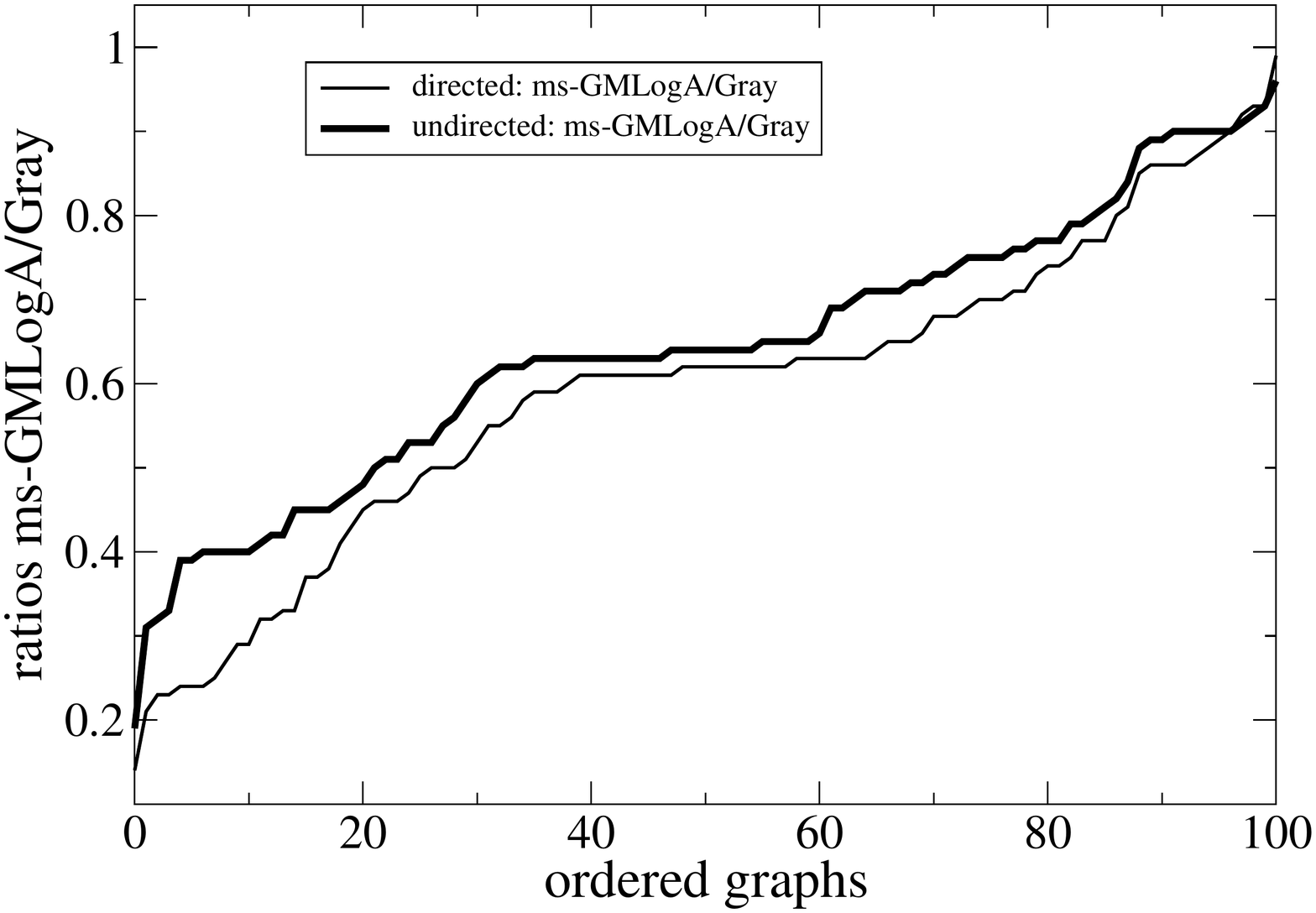}\\
(a) Gray ordering versus ms-GMLogA
\end{minipage}
\hspace{0.5cm} 
\begin{minipage}[b]{0.5\linewidth}
\centering
\includegraphics[width=7cm]{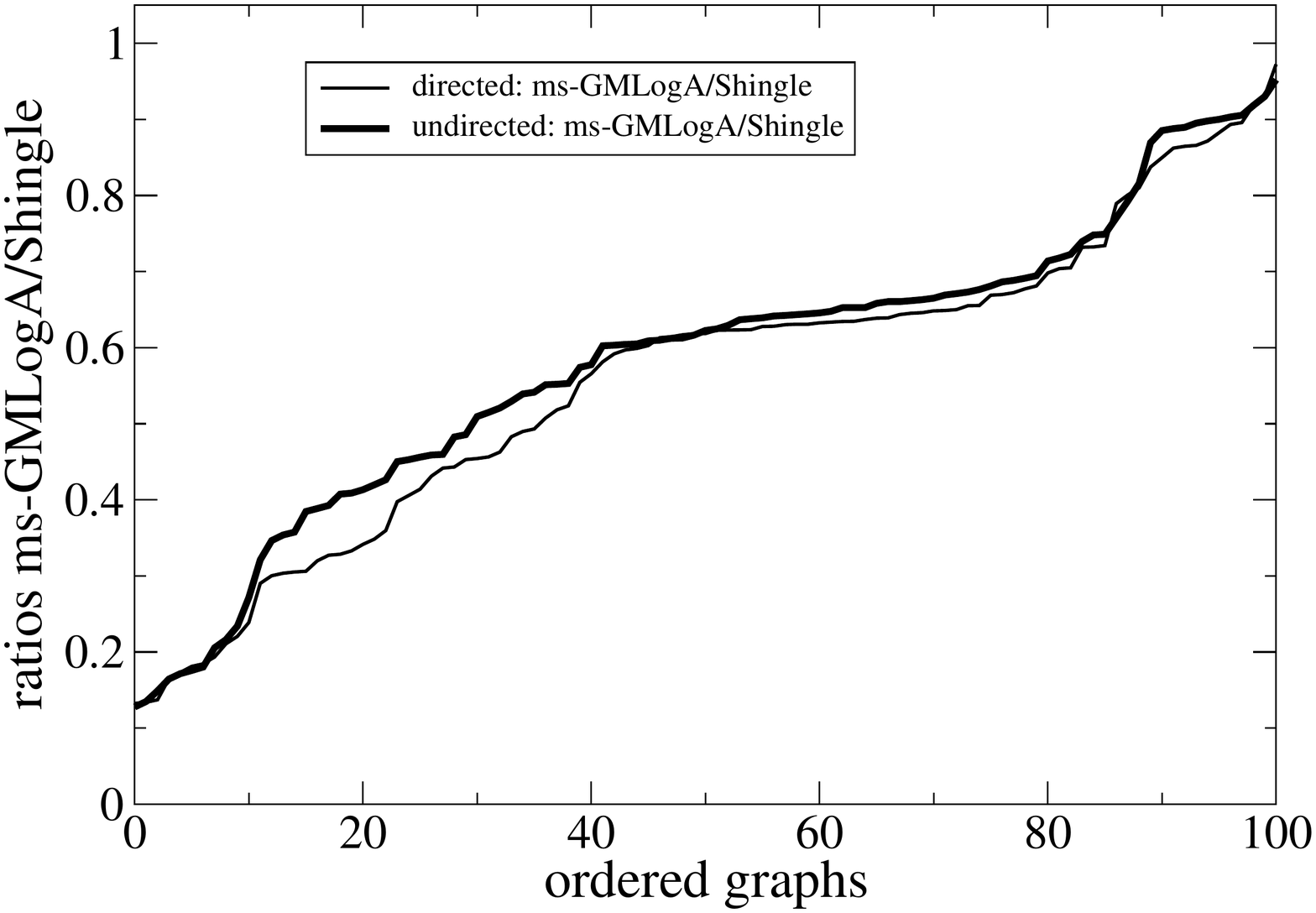}\\
(b) Double shingle versus ms-GMLogA
\end{minipage}
\caption{Comparison of Gray and double shingle orderings versus ms-GMLogA.}\label{fig:gray}
\end{figure}

 The Gray \cite{boldi-permuting} and double shingle \cite{compr-social} orderings have proven themselves as more successful heuristics than lexicographic and natural orderings \cite{mloga-site} and, thus, we present their comparison with ms-GMLogA in separate Figure \ref{fig:gray}(a-b). Two curves depicted in Figure \ref{fig:gray}(a) correspond to the comparison of Gray ordering on directed (regular curve) and undirected (bold curve) graphs, respectively. Each point on the curves corresponds to the ratio between $\beta$ values of ms-GMLogA with default parameters and Gray ordering, respectively. Similarly, the comparison of double shingle ordering \cite{compr-social} is shown in Figure \ref{fig:gray}(b). The (double) shingle heuristic is based on the node similarity ordering derived from estimation of Jaccard coefficients. In \cite{compr-social}, the authors prove that (double) shingle heuristic will be beneficial for networks that can be described by the preferential attachment model.
\par
\begin{figure}
\centering
\includegraphics[width=7cm]{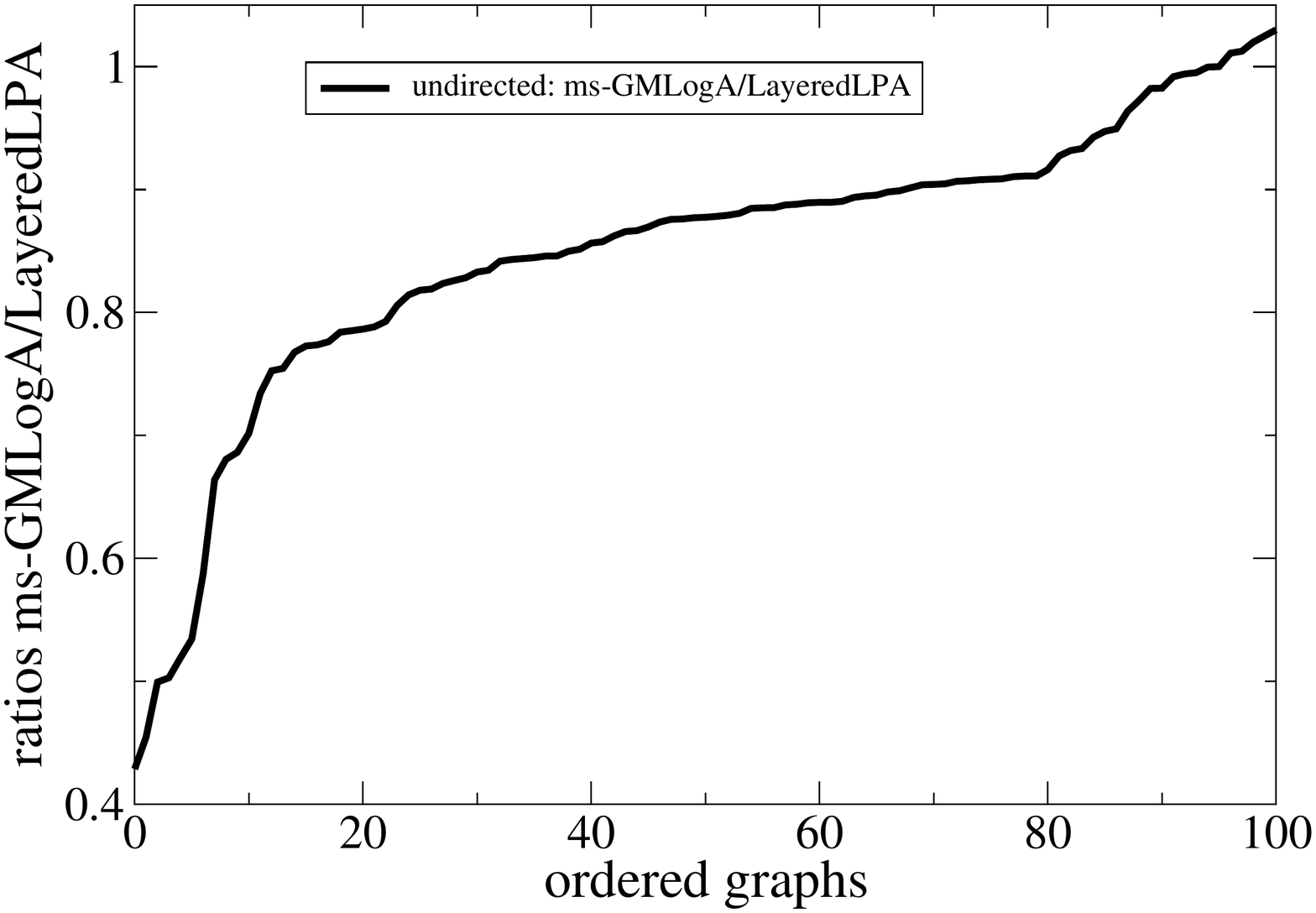}\\
\caption{Comparison of LayeredLPA versus ms-GMLogA.}\label{fig:potts}
\end{figure}

 Recently, the layered label propagation algorithm (LayeredLPA) has been introduced in \cite{webgr:impl}. This propagation method is based on the Potts model \cite{potts}. This algorithm is significantly more successful than natural, random, lexicographic, Gray and (double) shingle orderings. The success of the label propagation methods has a similar nature to the success of the algebraic distance coupling (see Section \ref{sec:algdist} and \cite{chen-safro-algdist-full}) in which the propagation and averaging of random values over the node neighborhoods is employed. However, our multiscale method shows better results (see Figure \ref{fig:potts}). Note that the LayeredLPA is introduced for undirected graphs only. We believe, that introducing the AMG-based framework to the label propagation model can significantly improve its quality.

\subsection{Scalability of ms-GMLogA}
\par
\begin{figure}
\begin{center}
\includegraphics[width=10cm]{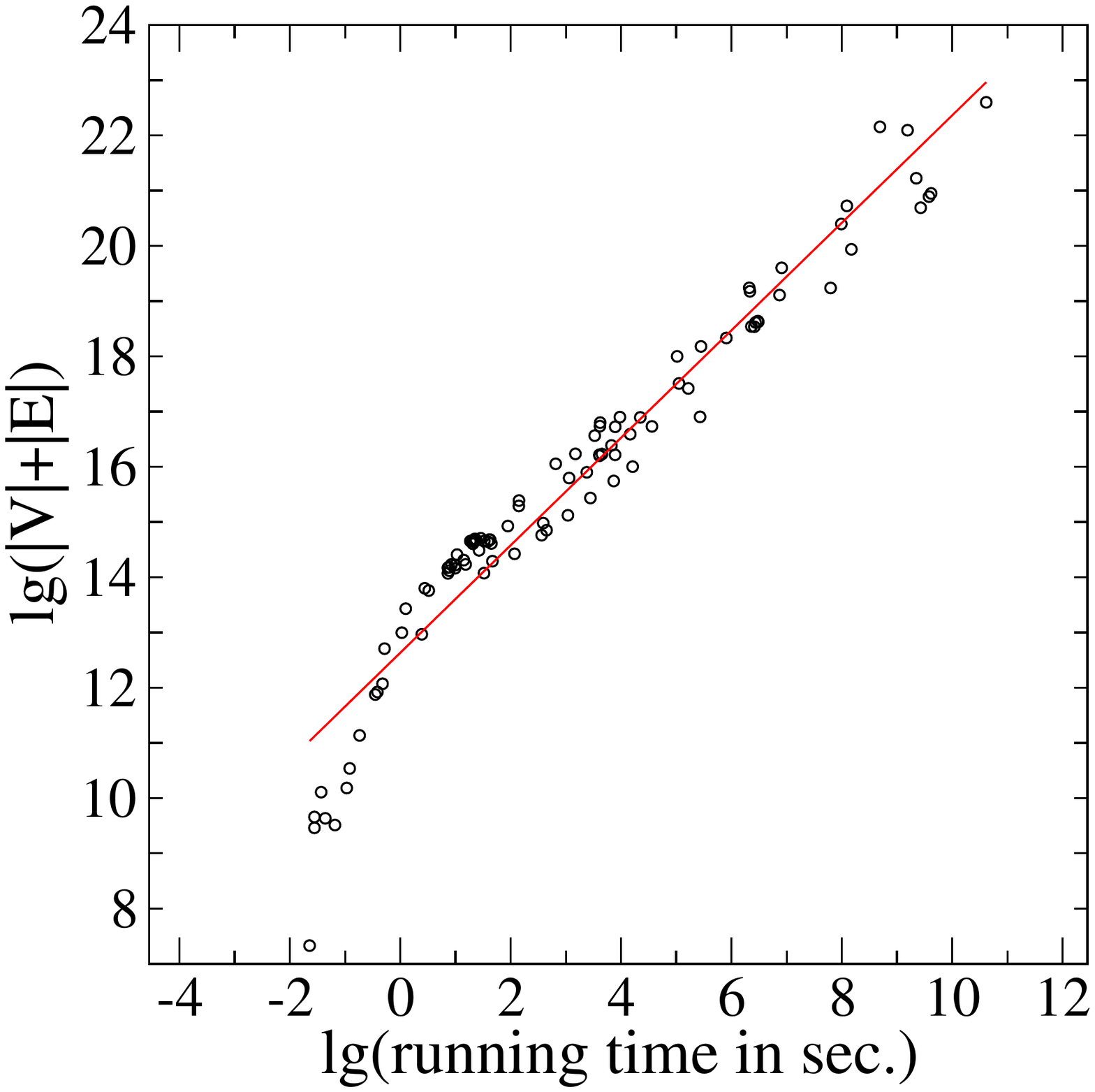}
\caption{ms-GMLogA: running time vs graph size.}\label{fig:scal}
\end{center}
\end{figure}

 Good scalability is one of the most important advantages of the multiscale algorithms. All components of our scheme are of linear complexity, and the total complexity is $O(|V|+|E|)$. The dependence of the running time on the graph size is depicted in Figure \ref{fig:scal}. Each small circle in the figure corresponds to a particular graph. We added to this figure a regression line whose slope is close to 1. The dependence is presented in logarithmic scale. We do not expect a more precise relationship between $|V| + |E|$ and the running time since the
implementation is far from being optimized and since it may depend also on structural factors such as the degree distribution of the  graph and its diameter.

\subsection{Spectral approach}
\par Arrangement of the graph vertices according to the eigenvector (called the Fiedler vector) corresponding to the second smallest eigenvalue of the graph Laplacian is a well-known and successful heuristic used for many ordering problems such as the minimum $p$-sum and the minimum envelope reduction (see \cite{pothen-envelope,juvan}). Usually, for large graphs the computation of the Fiedler vector is too expensive, and some approximation is employed \cite{Saad:1992:NML}. Most of these approximations can be viewed as a global averaging process that works until a particular convergence. As a result, during this process the vertex tends to be located at the weighted average of its neighbors, which makes it suitable for solving the quadratic functionals such as the minimum $2$-sum problem \cite{juvan} but creates a poor solution for the sum of logarithms. Several examples (from \cite{snap}) of a comparison between spectral approach and ms-GMLogA are presented in Table \ref{tab:spectral}.
 \begin{table}
\begin{center}
\caption{Comparison of spectral method and ms-GMLogA}\label{tab:spectral}
 \begin{tabular}{|l|c|c|}
 \hline
Network & Spectral &  ms-GMLogA\\
\hline
 as-caida20071105 & 12.4471 & 6.32\\
 email-Enron &  10.3117 &  7.06\\
 oregon1\_010407 &  11.0662  & 6.18\\
 p2p-Gnutella06 &  10.6513 &  7.13\\
 wiki-Vote &  9.2152 &  7.41\\
 \hline
 \end{tabular}
\end{center}
 \end{table}

\subsection{Compressing the ordered networks}
\input{bpl}

\section{Conclusions and future work}\label{sec:concl}
\par We have proposed a fast linear algorithm (ms-GMLogA), for compression-friendly graph reordering. The algorithm belongs to the family of multiscale methods \cite{vlsicad}. It uses a novel AMG-based coarsening scheme that is reinforced by a modification of algebraic distance couplings \cite{safro2010}. The empirical results on a large benchmark demonstrate its quality and scalability. The model takes into account the edge and node weights that can express different properties of a network model such as known link importance or its access frequency. Overall, we recommend this multiscale  framework for practical, compression-friendly network  orderings.
\par As future work we identify the following research directions that have the potential to improve ms-GMLogA:
\begin{itemize}
\item development of a learning algorithm to improve the bandwidth $h$ in (\ref{eq:dens});
\item development of a more sophisticated collective refinement of nodes;
\item development of a multiscale method for MLinGapA (defined below); and
\item more sophisticated multiscale organization of directed graphs.
\end{itemize}
\par A more sophisticated collective refinement of nodes can allow interruption of the coarsening earlier. Thus, if some more qualitative approximation can be achieved at the very coarse levels instead of solving exactly the coarsest and applying fast simple refinement at the next few finer level, it can potentially lead to better solutions at the fine levels.
\par The next natural step is to design first a refinement and then a full multiscale approach for {\it the minimum logarithmic gap arrangement problem} (MLinGapA) defined in \cite{compr-social} as an alternative model for compression. In this problem the goal is to find a permutation of nodes that minimizes
\begin{equation}
\sum_{i\in V} f_{\pi}(i,out(i)), 
\end{equation}
where $f_{\pi}(i,out(i))$ is a sum of logarithms of gaps between consecutive neighbors of $i$ ordered by $\pi$. This problem is also NP-hard \cite{compr-social}; however, we believe that having a suitable refinement and relaxation algorithms would make it possible to adopt the multiscale framework for this problem as well.
\par The last  research direction is related to finding a better multiscale scheme for directed graphs. The presented model constructs an undirected edge $ij$ from a given set of directed edges between $i$ and $j$ by reflecting both directions as an edge weight. In fact, this way of representation is not far from the attempts to solve the problems on nonsymmetric matrix $A$ by applying the known techniques on $A+A'$ or $A\cdot A'$. It is known that these methods suffer from several drawbacks, however; thus, we identify a finding of a more advanced multiscale representation for directed graphs as one of the major future research directions.

\section*{Acknowledgments}
This work was partially funded by the CSCAPES institute, a DOE project, and in part by DOE Contract DE-AC02-06CH11357. We express our gratitude to Marco Rosa for helping to design experiments with Webgraph package \cite{webgr:impl}.

\section*{Appendix A: Parameters}
{\bf $\Theta_{1,2}$}. The threshold parameters $\Theta_1$ and $\Theta_2$ from (\ref{eq:coarse-cond}) are responsible for a control of complexity of the coarse level problem and the quality of coarsening. It is important to check their robustness when designing a multiscale framework. The presented computational experiments have been executed with $\Theta_1=\Theta_2=1/2$, however, other values $.2\leq \Theta_1,\Theta_2\leq.8$ exhibit a good behavior of numerical results as well. Of course, the larger values increase the running time as the coarse graphs become bigger.

{\bf $\omega$}. Detailed discussion about choosing parameter $\omega$ for Jacobi over-relaxation on general graphs is discussed in \cite{chen-safro-algdist-full}. In all our computational experiments this parameter was 0.5.
\bibliographystyle{plain}
\bibliography{logsum-arxiv}
\vspace*{1cm}
\hspace*{1.5in}{\scriptsize\framebox{\parbox{2.4in}{
The submitted manuscript has been created in part by UChicago Argonne, LLC, Operator of Argonne National Laboratory (``Argonne'').  Argonne, a U.S. Department of Energy Office of Science laboratory, is operated under Contract No. DE-AC02-06CH11357.  The U.S. Government retains for itself, and others acting on its behalf, a paid-up nonexclusive, irrevocable worldwide license in said article to reproduce, prepare derivative works, distribute copies to the public, and perform publicly and display publicly, by or on behalf of the Government.
}}}
\end{document}

%% file: coarsening.tex
\par In the present work, coarsening is interpreted as a modified process of weighted aggregation reinforced by the algebraic distance couplings for logarithmic sum minimization. For a detailed discussion about the algebraic distance-based weighted aggregation and multiscale graph organization related to the general graph optimization problems, we refer the reader to \cite{safro2010}. We will briefly repeat its basic components for the completeness of the paper and will concentrate on the modifications related to Problem \ref{probdef}.
\subsubsection{Algebraic distance coupling}\label{sec:algdist}
\par Algebraic distance-based coupling is a measure of connectivity strength between two nodes connected by an edge \cite{safro2010,chen-safro-algdist-full}. Given the Laplacian of a graph, denoted by $L=D-W$, where $W$ is a weighted adjacency matrix of a graph and $D$ is the diagonal matrix with entries $d_{ii}=\sum_jw_{ij}$, we define an iteration matrix $H$ for Jacobi over-relaxation (JOR) as
\[
 H = (D/\omega)^{-1}((1/\omega-1)D + W_l+W_u)~,
\]
where $0\leq \omega \leq 1$ (see Appendix A) and $W_l$ and $W_u$ are the strict lower and upper triangular parts of $W$, respectively.
\begin{defn}
The algebraic distance coupling $\rho_{ij}$ is defined as
\[
\rho_{ij} = 1/\big( \sum_{r=1}^R \lg |\chi_i^{(k,r)} - \chi_j^{(k,r)}| \big)~,
\]
where $\chi^{(k,r)} = H^k \chi^{(0,r)}$ is a relaxed randomly initialized test vector (i.e., $\chi^{(0,r)}$ is a random vector sampled over \text{[-1/2, 1/2]}), R is a number of initial test vectors, and k is a number of iterations.
\end{defn}
Note that this definition is a modified version of the original definition from \cite{safro2010,chen-safro-algdist-full} which makes it more suitable for Problem \ref{probdef}. Several interesting properties (including convergence and model description) still can be proved similarly to \cite{chen-safro-algdist-full}.
\subsubsection{Coarse graph construction}
\par Considering $\rho_{ij}$ as an edge strength measure, one can construct a coarse graph by defining a classical AMG interpolation matrix $P$ that will project the fine graph to the coarse graph (i.e., to the lower-dimensional space). The projection is represented as 
\[
 L_c \leftarrow P L_f P^T~,
\]
where $L_f$ and $L_c$ are the Laplacians of fine ($G_f$) and coarse ($G_c$) graphs, respectively.
\par We begin by selecting a set of seed nodes $C\subset V_f$ that will represent the centers of future coarse nodes (or aggregates). In fact, $C$ is interpreted as a dominating set of $V_f$ (not necessarily the minimum size) such that other (fine) nodes $F = V_f \setminus C$ should be strongly coupled to $C$. This can be done by traversing all nodes and identifying for every visited node $i$ whether 
\begin{equation}\label{eq:coarse-cond}
 \sum_{j\in C} \rho_{ij}/ \sum_{j\in V_f} \rho_{ij} \geq \Theta_1 ~~~ \text{ and } ~~~  \sum_{j\in C} w_{ij}/ \sum_{j\in V_f} w_{ij} \geq \Theta_2~,
\end{equation}
where $\Theta_{(1,2)}$ are the parameters of coupling strength (see Appendix A). The order in which the nodes are traversed is based on the {\it future volume} principle \cite{safro2010}, which is a measure of how large (representative in terms of the current minimization problem) a coarse node can be. To keep the linear complexity of the entire framework, the order need not be calculated exactly but only roughly (for example using bucketing sort). In contrast to the multiscale approach for the generalized minimum $p$-sum problem \cite{safro2010}, we found that automatically moving to $C$ those nodes that have exceptionally large future volume is not necessary and even has a small negative impact in several social networks.
\par After identifying $C$, we define for each $i\in F$ its coarse neighborhood $N^c_i$ that contains a limited set of $C$-nodes to which $i$ is connected. The criterion for choosing $C$-nodes to $N^c_i$ is also based on $\rho_{ij}$. Let $I(j)$ be the ordinal number in the coarse graph of the node that represents the aggregate around a seed whose ordinal number at the fine level is $j$. The classical AMG interpolation matrix $P$ is defined by
\begin{equation}\label{interpmat}
P_{iI(j)}~=~ \left\{
  \begin{array}{lll}
  w_{ij}/\sum\limits_{k\in N^c_i}w_{ik} & \textsf{for }i\in F , ~ j\in N^c_i \\
  1               & \textsf{for }i\in C, ~ j=i \\
  0               & \textsf{otherwise} ~~~.
  \end{array} \right.
\end{equation}
$P_{iI(j)}$ thus represents the likelihood of $i$ belonging to the $I(j)$th aggregate. The edge connecting two coarse aggregates $p$ and $q$ is assigned with the weight $w_{pq}=\sum_{k\not= l}P_{kp}w_{kl}P_{lq}$. The volume of the $p$th coarse aggregate is $\sum_j v_jP_{jp}$. 

%% file: one_node.tex
\par Before proceeding to the stages of coarse-to-fine projection of a coarse-level solution, we describe how to approximate a solution of Problem \ref{probdef} for a single node only.
This will be a basic step in the initialization and relaxation described in Sections \ref{sec:init} and \ref{sec:relax}.

Denote by $N_i$ the set of $i$th neighbors with already assigned coordinates $\tilde{x}_j$.
To minimize the local contribution of $i$ to the total energy (\ref{probeq}), we assign to it a coordinate $x_i$  that minimizes
\begin{equation}\label{generalEnergy}
    \sum_{j\in N_i} w_{ij} \lg|x_i - \tilde{x}_j|~.
\end{equation}

Since for every $j \in N_i$, $x_i = \tilde{x}_j$ implies that the sum (\ref{generalEnergy}) is  minus infinity, we resolve this ambiguity by setting
\begin{equation}\label{eq:onenode}
  x_i = \tilde{x}_t \iff t = \arg\min_{k \in N_i} \sum_{k \neq j\in N_i} w_{kj} \lg|\tilde{x}_k - \tilde{x}_j|~.
\end{equation}
The trivial exact solution has a running time $O(|N_i|^2)$, as it requires to compute $|N_i|$ sums, each one with $|N_i|$ terms. Thus, to preserve the linear complexity of the entire algorithm, one can use the trivial solution for nodes with small $|N_i|$ only. We will approximate (\ref{eq:onenode}) using the heuristic that seeks the nearly minimum sum in the point of maximal density.
\par Consider set $\{\tilde{x}_j:~ j \in N_i\}$ as independent and  identically distributed  samples of a random variable with unknown distribution and $w_{ij}$ as the posteriori probability of a sample $\tilde{x}_j$. Assuming that, we have to choose a point where the estimated probability density is maximized. Various approaches exist for density estimation  \cite{silverman}. One of the most popular is called the ``Parzen window'' (or kernel density estimation) method. In this method, the  density at point $x$ is estimated as
\begin{equation}\label{eq:dens}
  d(x) = \frac{1}{|N_i|h}\sum_{j \in N_i} K\left(\frac{|x - \tilde{x}_j|}{h}\right)~,
\end{equation}
where $K$ is a kernel and $h$ is a smoothing parameter called the bandwidth.
To simplify calculations, we choose a kernel similar to the Gaussian kernel. Thus, the estimated density at point $x$ is
$$
  \hat{d}(x) = \frac{1}{|N_i|h}\sum_{j \in N_i} w_{ij} 2^{|x - \tilde{x}_j|/h}~.
$$
Note that the factor $\frac{1}{|N_i|h}$ may be omitted without changing the maximal's location.
Moreover, it is sufficient to calculate $\hat{d}(\tilde{x}_j)$ for every $j \in N_i$.
This approach allows us to compute the density in all $|N_i|$ points in linear time using following algorithm.

\begin{algorithm}
  \caption{Linear algorithm for density estimation}
  \label{linDensityAlgo}
  \KwIn{$x_1, \ldots, x_k$ - samples in increasing order}
  \KwIn{$h$ - smoothing parameter (window)}
  \KwIn{$p_1, \ldots, p_k$ - posteriory probabilities}
  \KwOut{$\hat{d}_1, \ldots, \hat{d}_k$ - estimated density}

  \tcc{calculate incrementally influence of left neighbors}
  $s_1 \gets p_1$

  \For {$t = 2$ to $k$} {
    $s_t = p_t + s_{t - 1}2^{(x_{t - 1} - x_t)/h}$
  }
  \tcc{calculate incrementally influence of right neighbors}
  $r_k \gets p_k$

  \For {$t = k - 1$ to $1$} {
    $r_t = p_t + s_{t + 1}2^{(x_{t} - x_{t + 1})/h}$
  }

  \tcc{aggregate results}
  \For {$t = 1$ to $k$} {
    $\hat{d}_t = s_t + r_t - p_t$
  }
\end{algorithm}

Likewise, for all density estimations techniques the crucial step is to choose the bandwidth $h$. On the one hand, it should be big enough to smooth the peaks and highly-oscillatory components. On the other hand, if $h$ is too big, then the maximum density will always be located at the middle. Based on experiments, we choose $h = N/2\lg(N)$, where $N = \max(\tilde{x}_i) - \min(\tilde{x}_i)$.

\begin{figure}
\centering
\includegraphics[width=10cm]{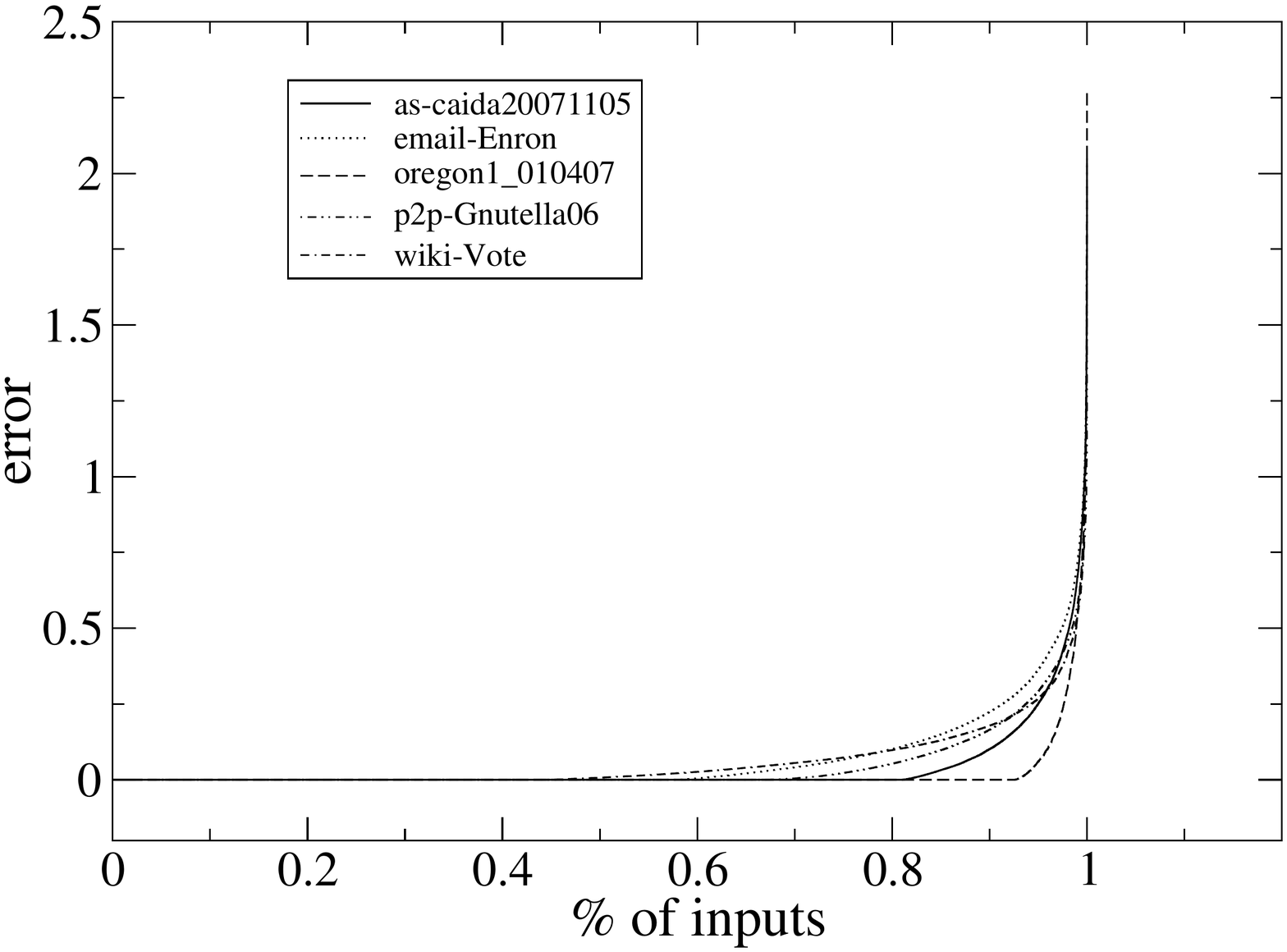}\\
\caption{Error distribution for density approximation on real-life networks. Each curve corresponds to one network. Each point on a curve corresponds to one difference between the exact (quadratic) solution of \ref{eq:onenode} and its approximation. The comparisons (points) are ordered by their errors in an ascending order.}\label{fig:babushka}
\end{figure}

To justify the chosen heuristic, we present our results from experiments on real data. In Figure \ref{fig:babushka} we show the error distribution for our approximation. The error is calculated as
$$
  \text{error} = \hat\theta - \theta^*~,
$$
where $\theta^*$ is the energy (excluding $-\infty$ terms) at the exact solution of (\ref{eq:onenode}) and $\hat\theta$ is an approximation. We can easily see that most of the time the approximation error is either zero or very small.


%% file: bpl.tex
\begin{figure}
\begin{minipage}[b]{0.5\linewidth} 
\centering
\includegraphics[width=7cm]{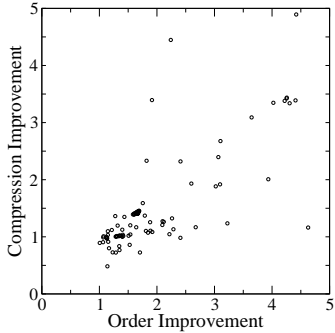} \\
a) initial ordering is natural
\end{minipage}
\begin{minipage}[b]{0.5\linewidth} 
\centering
\includegraphics[width=7cm]{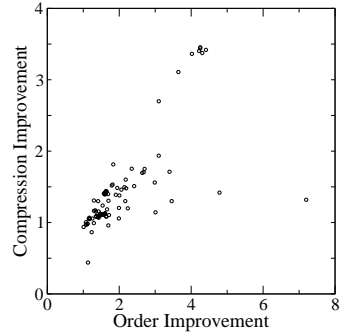} \\
b) initial ordering is Gray
\end{minipage}
\caption{Correlation between ordering and compression improvements}\label{fig:ordervscor}
\end{figure}

In \cite{compr-social}, Chierichetti et al. conjectured that minimizing the unweighted version of MLogA will automaticaly improve the graph compression. The conjecture has been supported by experimental observations on few web graphs and social networks.
However, for general networks, a better ordering does not always imply a better compression ratio produced by the kind of compressions described in \cite{boldi-vigna}. To illustrate that we present in Figure \ref{fig:ordervscor} a dependence of order improvement on the compression produced in \cite{boldi-vigna}.  \par Given an initial ordering (natural or Gray), we apply our algorithm to obtain a new ordering with smaller $\beta$. The x-axis corresponds to the ratio  between $\beta$ values of initial ordering and ms-GMLogA. The y-axis corresponds to the ratio obtained by compressing these orders. On most of the graphs, the conjecture is confirmed. However, on some graphs we can see the degradation of the compression results. It is observable, in particular, in Figure \ref{fig:ordervscor}(b), where the initial arrangement was obtained by Gray ordering that is significantly better than the natural ordering.

%% file: paper-arxiv.bbl
\begin{thebibliography}{10}

\bibitem{webgr:impl}
Laboratory for {W}eb {A}lgorithmics.
\newblock http://law.dsi.unimi.it/.

\bibitem{micah-compr}
Micah Adler and Michael Mitzenmacher.
\newblock Towards compressing web graphs.
\newblock In {\em DCC '01: Proceedings of the Data Compression Conference},
  page 203, Washington, DC, USA, 2001. IEEE Computer Society.

\bibitem{ad-gcbfs-09}
Alberto Apostolico and Guido Drovandi.
\newblock Graph compression by {B}{F}{S}.
\newblock {\em Algorithms}, 2(3):1031--1044, 2009.

\bibitem{pothen-envelope}
S.T. Barnard, A.~Pothen, and H.~Simon.
\newblock A spectral algorithm for envelope reduction of sparse matrices.
\newblock {\em Numerical Linear Algebra with Applications}, 2(4):317--334,
  1995.

\bibitem{boldi-vigna}
P.~Boldi and S.~Vigna.
\newblock The webgraph framework {I}: compression techniques.
\newblock In {\em WWW '04: Proceedings of the 13th {I}nternational {C}onference
  on World Wide Web}, pages 595--602, New York, NY, USA, 2004. ACM.

\bibitem{boldi-permuting}
Paolo Boldi, Massimo Santini, and Sebastiano Vigna.
\newblock Permuting web graphs.
\newblock In Konstantin Avrachenkov, Debora Donato, and Nelly Litvak, editors,
  {\em WAW}, volume 5427 of {\em Lecture Notes in Computer Science}, pages
  116--126. Springer, 2009.

\bibitem{boman-advances}
Erik~G Boman, Umit~V Catalyurek, CÃ©edric Chevalier, Karen~D Devine, Ilya
  Safro, and Michael~M Wolf.
\newblock Advances in parallel partitioning, load balancing and matrix ordering
  for scientific computing.
\newblock {\em Journal of Physics: Conference Series}, 180(1), 2009.

\bibitem{bmr}
A.~Brandt, S.~McCormick, and J.~Ruge.
\newblock Algebraic multigrid ({AMG}) for automatic multigrid solution with
  application to geodetic computations.
\newblock Technical report, Institute for Computational Studies, Fort Collins,
  CO, POB 1852, 1982.

\bibitem{vlsicad}
A.~Brandt and D.~Ron.
\newblock Chapter 1 : Multigrid solvers and multilevel optimization strategies.
\newblock In J.~Cong and J.~R. Shinnerl, editors, {\em Multilevel Optimization
  and VLSICAD}. Kluwer, 2003.

\bibitem{chen-safro-algdist-full}
Jie Chen and Ilya Safro.
\newblock Algebraic distance on graphs.
\newblock Technical Report ANL/MCS-P1696-1009 (Preprint), Argonne National
  Laboratory.

\bibitem{compr-social}
Flavio Chierichetti, Ravi Kumar, Silvio Lattanzi, Michael Mitzenmacher,
  Alessandro Panconesi, and Prabhakar Raghavan.
\newblock On compressing social networks.
\newblock In {\em KDD '09: Proceedings of the 15th ACM SIGKDD international
  conference on Knowledge discovery and data mining}, pages 219--228, New York,
  NY, USA, 2009. ACM.

\bibitem{wojciech-compress}
Yongwook Choi and Wojciech Szpankowski.
\newblock Compression of graphical structures.
\newblock In {\em ISIT'09: Proceedings of the 2009 IEEE international
  conference on Symposium on Information Theory}, pages 364--368, Piscataway,
  NJ, USA, 2009. IEEE Press.

\bibitem{davis}
T.~Davis.
\newblock University of {F}lorida {S}parse {M}atrix {C}ollection.
\newblock {\em NA Digest}, 97(23), 1997.

\bibitem{survey:petit}
J.~D\'iaz, J.~Petit, and M.~Serna.
\newblock A survey of graph layout problems.
\newblock {\em ACM Comput. Surv.}, 34(3):313--356, 2002.

\bibitem{juvan}
Martin Juvan and Bojan Mohar.
\newblock Optimal linear labelings and eigenvalues of graphs.
\newblock {\em Discrete Appl. Math.}, 36(2):153--168, 1992.

\bibitem{survey:lai}
Y.~Lai and K.~Williams.
\newblock A survey of solved problems and applications on bandwidth, edgesum,
  and profile of graphs.
\newblock {\em J. Graph Theory}, 31:75--94, 1999.

\bibitem{snap}
J.~Leskovec.
\newblock Stanford {N}etwork {A}nalysis {P}ackage ({S}{N}{A}{P}).
\newblock http://snap.stanford.edu/index.html.

\bibitem{soc-net-evol}
Jure Leskovec, Lars Backstrom, Ravi Kumar, and Andrew Tomkins.
\newblock Microscopic evolution of social networks.
\newblock In {\em KDD '08: Proceeding of the 14th ACM SIGKDD {I}nternational
  {C}onference on {K}nowledge {D}iscovery and {D}ata {M}ining}, pages 462--470,
  New York, NY, USA, 2008. ACM.

\bibitem{potts}
Renfrey~B. Potts.
\newblock Some generalized order-disorder transformations.
\newblock {\em Proceedings of the Cambridge Philosophical Society},
  48:106--109, 1952.

\bibitem{safro2010}
Dorit Ron, Ilya Safro, and Achi Brandt.
\newblock Relaxation-based coarsening and multiscale graph organization.
\newblock Technical Report ANL/MCS-P1741-0410 (Preprint), Argonne National
  Laboratory, 2010.

\bibitem{Saad:1992:NML}
Y.~Saad.
\newblock {\em Numerical Methods for Large Eigenvalue Problems}.
\newblock Halsted Press, 1992.

\bibitem{safro2003}
I.~Safro, D.~Ron, and A.~Brandt.
\newblock Graph minimum linear arrangement by multilevel weighted edge
  contractions.
\newblock {\em Journal of Algorithms}, 60(1):24--41, 2006.

\bibitem{safro2005}
I.~Safro, D.~Ron, and A.~Brandt.
\newblock Multilevel algorithms for linear ordering problems.
\newblock {\em Journal of Experimental Algorithmics}, 13:1.4--1.20, 2008.

\bibitem{mloga-site}
I.~Safro and B.~Temkin.
\newblock Benchmark for the {M}inimum {L}ogarithmic {A}rrangement {P}roblem.
\newblock http://www.mcs.anl.gov/~safro/mloga.html.

\bibitem{silverman}
B.~W. Silverman.
\newblock {\em Density Estimation for Statistics and Data Analysis}.
\newblock Chapman \& Hall, London, 1986.

\bibitem{bollt-compr}
Jie Sun, Erik~M Bollt, and Daniel Ben-Avraham.
\newblock Graph compression-save information by exploiting redundancy.
\newblock {\em Journal of Statistical Mechanics: Theory and Experiment},
  2008(06):P06001, 2008.

\bibitem{walshaw-inbook}
Chris Walshaw.
\newblock {\em Multilevel Refinement for Combinatorial Optimisation: Boosting
  Metaheuristic Performance}, pages 4261--289.
\newblock Springer, 2008.

\bibitem{soc-locality}
H.~Zhang, B.~Qiu, K.~Ivanova, C.~L. Giles, H.~C. Foley, and J.~Yen.
\newblock Locality and attachedness-based temporal social network growth
  dynamics analysis.
\newblock {\em Journal of American Society of Information Science and
  Technology}, 61:964--977, 2009.

\end{thebibliography}
